\documentclass[12pt]{article}
\usepackage{cite}
\usepackage{enumerate}
\usepackage{ifpdf}

\usepackage{ae}
\usepackage[T1]{fontenc}
\usepackage[ansinew]{inputenc}
\usepackage{csquotes}
\usepackage{amsmath}
\usepackage{relsize}
\usepackage{amssymb}
\usepackage{tikz}
\usepackage{graphicx}
\usepackage{color}
\definecolor{darkblue}{cmyk}{0.9,0.9,0,0}
\definecolor{darkgreen}{rgb}{0,0.55,0}
\usepackage[colorlinks=true,linkcolor=darkblue,citecolor=darkblue,urlcolor=darkblue]{hyperref}
\usepackage{epsfig}
\usepackage{graphicx}
\usepackage[bulletsep]{collref}

\newcommand{\beq}{\begin{equation}}
\newcommand{\eeq}{\end{equation}}
\newcommand{\beqy} {\begin{eqnarray}}
\newcommand{\eeqy} {\end{eqnarray}}
\newcommand{\bsmat}{\begin{smallmatrix}}
\newcommand{\esmat}{\end{smallmatrix}}
\newcommand{\bmat}{\begin{matrix}}
\newcommand{\emat}{\end{matrix}}
\newcommand{\sfrac}[2]{{\textstyle\frac{#1}{#2}}}

\def\({\left(}
\def\){\right)}
\def\[{\left[}
\def\]{\right]}

\def\<{\langle}
\def\>{\rangle}

\usepackage{varioref}
\usepackage{makeidx}
\makeindex

\usepackage[english]{babel}
\usepackage{subfig}

\usepackage{tabularx}
\begin{document}

\thispagestyle{empty}

\renewcommand{\thefootnote}{\fnsymbol{footnote}}
\setcounter{page}{1}
\setcounter{footnote}{0}
\setcounter{figure}{0}

\begin{titlepage}

\begin{center}

\vskip 2.3 cm 

\vskip 5mm

{\Large \bf Large spin systematics in CFT}
\vskip 0.5cm

\vskip 15mm

{Luis F. Alday, Agnese Bissi and Tomasz {\L}ukowski}

\vspace{1cm}
\centerline{{\it Mathematical Institute, University of Oxford,}}
\centerline{{\it Andrew Wiles Building, Radcliffe Observatory Quarter,}}
\centerline{{\it Woodstock Road, Oxford, OX2 6GG, UK}}

\end{center}

\vskip 2 cm

\begin{abstract}
\noindent Using conformal field theory (CFT) arguments we derive an infinite number of constraints on the large spin expansion of the anomalous dimensions and structure constants of higher spin operators. These arguments rely only on analyticity, unitarity, crossing-symmetry and the structure of the conformal partial wave expansion.  We obtain results for both, perturbative CFT to all order in the perturbation parameter, as well as non-perturbatively. For the case of conformal gauge theories this provides a proof of the reciprocity principle to all orders in perturbation theory and provides a new "reciprocity" principle for structure constants. We argue that these results extend also to non-conformal theories. 
\end{abstract}

\end{titlepage}

\setcounter{page}{1}
\renewcommand{\thefootnote}{\arabic{footnote}}
\setcounter{footnote}{0}

\section{Introduction}
In the last few years there has been an increasing interest in conformal field theories (CFT) in dimensions higher than two. Much of this interest is due to the effectiveness of the conformal bootstrap program \cite{Rattazzi:2008pe}. A CFT is characterized by the spectrum of scaling dimensions of all its primary operators, together with the structure constant for any three given primaries. The structure of the operator product expansion (OPE) then allows to write any correlation function, given this {\it CFT data}. The conformal bootstrap program consists in constraining the {\it CFT data} by requiring consistency of higher correlation functions, {\it e.g.} crossing-symmetry, together with basic properties of well behaved CFT's, such as unitarity and the structure of the OPE.  

A particularly interesting class of operators in any theory (not necessarily conformal) are operators with high spin. It has been argued \cite{Gross:1974cs,Georgi:1951sr,Korchemsky:1988si} that in gauge theories the scaling dimension shows a logarithmic behavior 
\begin{equation}
\label{larges}
\Delta_\ell-\ell= \Gamma(g) \log \ell+\ldots
\end{equation}
for large spin $\ell$. We will loosely denote these operators as "single-trace" operators. The function $\Gamma(g)$ is called the cusp anomalous dimension and is of great interest, since it makes its appearance in various contexts. 

One of the questions we will address in this paper is the systematic expansion of the scaling dimension for single-trace operators in inverse powers of the spin. It has been observed that higher terms in the $1/\ell$ expansion, for single-trace operators, satisfy the so called {\it reciprocity} principle. The reciprocity principle is most easily described for conformal field theories.  Suppose the twist of the operators at tree-level is two so that $\Delta_\ell-\ell=2+ \gamma_\ell$. Conformal symmetry implies that the anomalous dimension is actually a function of the conformal spin \cite{Basso:2006nk}
\begin{equation}
\gamma_\ell = f(\ell+\sfrac{1}{2}\gamma_\ell).
\end{equation}
Reciprocity is equivalent to the statement that when expanded in inverse powers of the bare Casimir $J^2_b=\ell(\ell+1)$, $f(\ell)$ contains only even powers of $J_b$. This parity preserved property of $f(\ell)$ was originally proposed in \cite{Dokshitzer:2005bf} and extensively checked in \cite{Basso:2006nk}, for several examples and at high loops in perturbation theory. For our purposes it is convenient to phrase the reciprocity principle as follows. Given the anomalous dimension we define the Casimir
\begin{equation}
J^2 =(\ell+\gamma_\ell/2)(\ell+1+\gamma_\ell/2)\,.
\end{equation}
Then we can consider the expansion of $\gamma_\ell$ for large values of $J$. The reciprocity principle as stated above is equivalent to the expansion taking the form
\begin{equation}
\gamma_\ell = \alpha_0(\log J)+ \frac{\alpha_1(\log J)}{J^2}+  \frac{\alpha_2(\log J)}{J^4}+\ldots
\end{equation}
where odd powers of $J$ are absent. 

Another class of operators with high spin corresponds to what we will call double-trace operators. In \cite{Fitzpatrick:2012yx,Komargodski:2012ek} it has been shown that in any CFT, given a scalar primary field $\phi$, there are infinite towers of operators with dimension
\begin{equation}
\label{rec}
2\Delta_\phi +2n -\frac{c}{\ell^{\tau_{min}}}+\ldots
\end{equation}
where $\tau_{min}$ is the twist of the minimal twist operator appearing in the OPE of $\phi$ with itself (besides the identity operator). Usually one can exchange the stress tensor operator. In this case $\tau_{min}=2$. 

In \cite{Alday:2007mf} a geometrical argument based on conformal symmetry, was given for both, the logarithmic behavior of single-trace operators (\ref{larges}) as well as the behavior (\ref{rec}) for double-trace operators. In this paper we will be concerned with the higher order terms in the $1/\ell$ expansion. The systematic approach of  \cite{Fitzpatrick:2012yx,Komargodski:2012ek}, which is based on crossing-symmetry of correlators, is more suitable for this purpose. 

In this paper we will consider correlators of identical operators which contain higher spin operators as intermediate states. We will study how the expansion in inverse powers of the spin is constrained by conformal symmetry, crossing symmetry, unitarity and analyticity of the full correlator. In section two we will restrict our attention to perturbative CFT. Our main result is a proof, to arbitrary loop order, of the reciprocity principle for the anomalous dimension of single-trace leading twist operators with higher spin. Furthermore, we derive a new set of infinite relations involving the expansion of the structure constants. We comment on application to non-conformal theories. In section three we consider  a non-perturbative CFT and derive analogous results for the anomalous dimension of double-trace operators with higher spin. These results include the next order correction to the results of \cite{Fitzpatrick:2012yx,Komargodski:2012ek}, but in addition imply conditions for the higher order terms. We end up with some discussion. In the appendices we apply our method to interesting cases not covered in the body of the paper. 

\section{Large spin systematics for perturbative CFT}

\subsection{Results}

We start by describing our method in the context of perturbative CFT. Consider for definiteness the four-point function of four identical real scalar operators ${\cal O}$ of dimension $\Delta_{\cal O}$ in a four-dimensional CFT. Conformal invariance implies
\begin{equation}
\langle {\cal O}(x_1){\cal O}(x_2){\cal O}(x_3){\cal O}(x_3) \rangle = \frac{{\cal G}(u,v)}{x_{12}^{2\Delta_{\cal O}}x_{34}^{2\Delta_{\cal O}}}
\end{equation}
where we have introduced the cross-ratios
\begin{equation}
u=\frac{x_{12}^2x_{34}^2}{x_{13}^2x_{24}^2} = z \bar z,~~~~~v=\frac{x_{14}^2x_{23}^2}{x_{13}^2x_{24}^2} = (1-z)(1-\bar z)
\end{equation}
and $z,\bar z$ have been introduced for later convenience. Crossing symmetry reads
\begin{equation}
\label{crossing}
v^{\Delta_{\cal O}} {\cal G}(u,v) = u^{\Delta_{\cal O}} {\cal G}(v,u) \,.
\end{equation}
The above correlator can be expanded in terms of conformal partial waves. Expanding along the $s-$channel we obtain
\begin{equation}
\label{OPEexp}
{\cal G}(u,v) = 1+\sum_{\substack{\Delta,\ell}} a_{\Delta,\ell} G_{\Delta,\ell}(u,v)\,.
\end{equation}
In four dimensions the conformal blocks are given by
\begin{align}
G_{\Delta,\ell}(u,v) &= u^{\frac{1}{2}(\Delta-\ell)} g_{\Delta,\ell}(u,v)\,,\\
 g_{\Delta,\ell}(z,\bar z) &= (-1)^\ell \frac{1}{z-\bar z} \left( k_{\Delta+\ell}(z) k_{\Delta-\ell-2}(\bar z) -k_{\Delta+\ell}(\bar z) k_{\Delta-\ell-2}(z)\right)\,,\\
k_\beta(z)&=z^{\beta/2+1}F_{\beta/2}(z),~~~~~~~F_{\beta/2}(z) =~_2F_1\left(\frac{\beta}{2},\frac{\beta}{2},\beta,z \right)\,.
\end{align}

In a perturbative CFT with coupling constant $g$, ${\cal G}(u,v)$ admits an expansion
\begin{equation}
{\cal G}(u,v) = {\cal G}^{(0)}(u,v) + g \, {\cal G}^{(1)}(u,v)+\ldots
\end{equation}
Among the intermediate primary operators appearing in the OPE ${\cal O} \times {\cal O}$, there is a tower of leading twist conformal primary operators with spin $\ell=0,2,\ldots$, which in perturbation theory have dimension:
\begin{eqnarray}
\Delta_\ell = \ell+\tau_{0} +\gamma_\ell\,.
\end{eqnarray}
Analyticity of the explicit tree-level answer in $u$ and $1-v$, together with the structure of the conformal block\footnote{The conformal blocks admit an expansions involving $u^{\tau_{0}/2}$ times integer powers of $u$ and $(1-v)$.},  requires $\tau_{0}$  to be an even number. The anomalous dimension $\gamma_\ell$ is a small parameter, but can be an arbitrary function of the coupling constant. Expanding ${\cal G}(u,v)$ in powers of $u$ it follows, from the explicit expression of the conformal blocks, that only the leading twist operators will contribute to the leading order. More precisely
\begin{eqnarray}
{\cal G}(u,v) = 1 + u^{\tau_{0}/2} h(\log u, v) +\ldots
\end{eqnarray} 
Considering the small $u$ limit of the conformal blocks we obtain the following decomposition for $h(\log u, v)$:
\begin{equation}
\label{OPEsmallu}
\sum_{\ell=0,2,\ldots} a_\ell u^{\tau_{0}/2+\gamma_\ell/2} (1-v)^\ell ~F_{\ell+\frac{\tau_{0}}{2}+\frac{\gamma_\ell}{2}}(1-v) = u^{\tau_{0}/2} h(\log u, v)\,,
\end{equation}
where the sum runs over the leading twist conformal primary operators. What can we say about the small $v$ behavior of $h(\log u, v)$?  The OPE structure for  free CFT was extensively studied in  \cite{Dolan:2000ut}. At any order in perturbation theory, crossing symmetry plus the structure of the conformal partial waves expansion, imply the small $v$ behavior of free theories, up to multiplication by powers of $\log v$. Hence, we expect 
 \begin{equation}
 \label{divergence}
 u^{\tau_{0}/2} h(\log u, v) \sim  u^{\tau_{0}/2} v^{-\tau_{0}/2}\,.
 \end{equation}
Here we are assuming leading twist operators of the schematic form $\varphi \partial^\ell \varphi$, where $\varphi$ is a real scalar field. In appendix (\ref{appnonscalar}) we consider more general cases and show that our conclusions remain unchanged. From (\ref{divergence}) we see $h(\log u, v)$ contains a divergence as $v$ becomes small. The only way to obtain such a divergence is by summing an infinite number of terms in (\ref{OPEsmallu}). Furthermore, the divergence will come solely from the region $\ell \gg 1$. In what follows the structure of higher powers of $v$ will be important. Note that analyticity of the tree-level result implies, at any order in perturbation theory, a structure of the form
\begin{equation}
h(\log u, v) = v^{-\tau_{0}/2}\left(h_0(\log u,\log v) + v h_1(\log u,\log v)+\ldots \right)\,,
\end{equation}
where only integer powers of $v$ appear. 

It is well known that the full conformal blocks are eigenfunctions of Casimir operators. The method we will use below relies on the existence of a Casimir operator  for the functions appearing in (\ref{OPEsmallu}):
\begin{equation}
f_{\ell+\tau_{0}+\gamma_\ell,\ell}(u,v) \equiv u^{\tau_{0}/2+\gamma_\ell/2} (1-v)^\ell ~F_{\ell+\frac{\tau_{0}}{2}+\frac{\gamma_\ell}{2}}(1-v)\,.
\end{equation}
More precisely, defining
\begin{equation}
\label{casimir}
{\cal D} = (1-v)^2 \partial_v -u(1-v)\partial_u + v(1-v)^2 \partial^2_v + v u^2 \partial_u^2 -2 u v(1-v)\partial_u \partial_v\,,
\end{equation}
we find 
\begin{equation}
{\cal D} f_{\ell+\tau_{0}+\gamma_\ell,\ell}(u,v)= J^2_{\ell+\tau_{0}+\gamma,\ell} f_{\ell+\tau_{0}+\gamma_\ell,\ell}(u,v)\,,
\end{equation}
where we have defined
\begin{equation}\label{casimir_pert}
J^2_{\ell+\tau_{0}+\gamma,\ell} \equiv  \frac{1}{4}(2\ell+\tau_{0}+\gamma_\ell)(2\ell+\tau_{0}+\gamma_\ell-2)\,.
\end{equation}
The operator ${\cal D}$ arises from considering the full Casimir operator in the small $u$ limit. Acting with ${\cal D}$ on the r.h.s. of $(\ref{OPEsmallu})$, increases by one the degree of divergence at small $v$. Consequently, on the l.h.s. of $(\ref{OPEsmallu})$ the behavior at large $\ell$ is enhanced. Furthermore, note that this operation does not spoil the property that only integer powers of $v$ appear in the small $v$ expansion. As we will see, reciprocity is a direct consequence of this very simple fact!

\bigskip

\noindent {\bf The method}

\bigskip

Let us consider the problem at tree-level. At this order $\gamma_\ell=0$ and $a_\ell = a_\ell^{(0)}$.  The structure of four-point functions for general CFT theories in the free-theory limit has been studied in \cite{Dolan:2000ut}. At tree-level the correlators can be simply computed by Wick contractions. For small $u$ we obtain \footnote{For instance, one can consider external operators of the form ${\cal O} = Tr \varphi^p$. In that case the leading-twist contribution comes from diagrams where operators one and two are connected by $p-1$ propagators.}
\begin{equation}
\sum_{\ell=0,2,\ldots} a_\ell^{(0)} u^{\tau_{0}/2} (1-v)^\ell ~F_{\ell+\frac{\tau_{0}}{2}}(1-v) \sim u^{\tau_{0}/2} \left( \frac{1}{v^{\tau_{0}/2}}+1\right)\,.
\end{equation}
In theories with a large central charge, the r.h.s. will be usually suppressed by a power of $c$, if the proper normalization is used. This factor will not play any role in our discussion, and our treatment will be valid for any value of the central charge. We can solve for $a_\ell^{(0)}$ and obtain
\begin{equation}
\label{atree}
a_\ell^{(0)}=\frac{2 \Gamma \left(\ell+\frac{\tau_0 }{2}\right)^2 \Gamma (\ell+\tau_0 -1)}{\Gamma (\ell+1) \Gamma \left(\frac{\tau_0 }{2}\right)^2 \Gamma (2 \ell+\tau_0 -1)}\,.
\end{equation}
 The large $\ell$ behavior of $a_\ell^{(0)}$ is fixed by the divergence in (\ref{divergence}). One could imagine non-generic CFT's with a different sub-leading behavior. This will not affect our subsequent discussion and our results will also apply to those. 

Let us now consider the sum (\ref{OPEsmallu}) in perturbation theory. We would like to evaluate the divergence of the sum as $v$ becomes small. A method to compute the leading divergence has been introduced in\cite{Alday:2007mf} and systematically developed  in  \cite{Fitzpatrick:2012yx,Komargodski:2012ek,Kaviraj:2015cxa}. First we introduce $v=\epsilon$, with the idea of expanding in powers of $\epsilon$. The divergence will come from the region of large $\ell$. We make this precise by introducing:
\begin{equation}
\label{measure}
\ell = \frac{x}{\epsilon^{1/2}},~~~~~\sum_\ell \rightarrow \frac{1}{2} \int_0^\infty dx\,.
\end{equation}
Furthermore, we introduce the integral representation for the hypergeometric function 
\begin{equation}
F_{\ell+\frac{\tau_{0}}{2}+\frac{\gamma_\ell}{2}}(1-v) = \frac{\Gamma(2\ell+\gamma_\ell+\tau_0)}{\Gamma(\ell+\gamma_\ell/2+\tau_0/2)^2} \int_0^1 \frac{(t(1-t))^{\ell-1+\gamma_\ell/2+\tau_0/2}}{(1-t(1-v))^{\ell+\gamma_\ell/2+\tau_0/2}}dt
\end{equation}
and perform the change of coordinates $t \rightarrow 1-t \epsilon^{1/2}$. This integral representation suggests we rescale the perturbative OPE coefficients as
\begin{equation}
\label{rescaledOPE}
a_\ell=\frac{2 \Gamma \left(\ell+\frac{\tau_0 }{2} + \frac{\gamma_\ell}{2} \right)^2 \Gamma (\ell+\tau_0+ \frac{\gamma_\ell}{2} -1)}{\Gamma (\ell+1+ \frac{\gamma_\ell}{2}) \Gamma \left(\frac{\tau_0 }{2}\right)^2 \Gamma (2 \ell+\tau_0+\gamma_\ell -1)} \hat a_\ell\,.
\end{equation}
Of course, at tree level $\hat a_\ell=1$. We will see these rescaled structure constants have definite reciprocity properties. In order to proceed, we perform a further change of variables and introduce the rescaled Casimir
\begin{equation}
\label{xtoj}
\frac{j^2}{\epsilon}=\left( \frac{x}{\epsilon^{1/2}}+\gamma_\ell/2\right)\left( \frac{x}{\epsilon^{1/2}}+1+\gamma_\ell/2\right)
\end{equation}
and then interpret the anomalous dimension and rescaled structure constants as a functions of $j$. As we will see below, this change of variables simplifies things drastically. Expanding the integrand in powers of $\epsilon$ we find the integral over $t$ is convergent and can be performed order by order, leading to 
\begin{align}
\label{smallvexpansion}
h(\log u,v)|_{v=\epsilon} = \epsilon^{-\tau_0/2}&\left( \frac{4}{\Gamma \left(\frac{\tau_0 }{2}\right)^2 } \int_0^\infty \hat a(j) u^{\gamma(j)/2} j^{\tau_0-1}K_0(2j)dj -\right. \\
&  \left.- \epsilon^{1/2} \frac{2}{\Gamma \left(\frac{\tau_0 }{2}\right)^2}  \int_0^\infty \hat a(j) u^{\gamma(j)/2} j^{\tau_0-1}K_0(2j) \gamma'(j) dj +\ldots \right) \nonumber\,.
\end{align}
Our claim is that this expansion reproduces not only the leading divergence in the small $v$ expansion, but actually all divergent terms! This is somewhat expected, since divergent terms (even subleading) do come from the tail in the sum over spins, but our claim is that the simple measure (\ref{measure}) does not receive corrections. Before proceeding, let us mention that this is valid regardless of the lowest bound $j_{0}$ in the integration region in (\ref{smallvexpansion}), provided $j_0$ is of order $\epsilon^{1/2}$. This is consistent with the fact that divergences come only from the tail in the sum over $\ell$ and we could have starting summing from any finite $\ell$.

In perturbation theory (but to an arbitrary loop order!) we expect $\gamma(j)$ and $\hat a(j)$ to have the following large $j$ expansion:
\begin{eqnarray}
\label{largejexpansions}
\gamma(j) = p_0(\log j^2/\epsilon) + \frac{ p_1(\log j^2/\epsilon) }{j} \epsilon^{1/2}+\frac{ p_2(\log j^2/\epsilon) }{j^2} \epsilon+\cdots\\
\hat a(j) = q_0(\log j^2/\epsilon) + \frac{ q_1(\log j^2/\epsilon) }{j} \epsilon^{1/2}+\frac{ q_2(\log j^2/\epsilon) }{j^2} \epsilon+\cdots \nonumber
\end{eqnarray}
where we have stressed the fact that the functions $p_i,q_i$ can depend logarithmically on $j$, but do not contain powers. Plugging these expansions into (\ref{smallvexpansion}) we obtain an integral expression for the small $v$ expansion of $h(\log u,v)$. Our claim implies that such expressions can be trusted provided the overall powers of $\epsilon$ are negative. On the other hand, remember that analyticity forbids non-integer powers of $v$, hence we obtain integral constraints on the functions $p_i$, $q_i$. For instance, absence of the leading half-integer power implies
\begin{equation}
\epsilon^{-\frac{\tau_0}{2}+\frac{1}{2}} \int_0^\infty j^{\tau_0-2} u^{\frac{1}{2} p_0} \left( 2 q_1 - q_0 p_0' + q_0 p_1 \log u \right)K_0(2j) dj = 0
\end{equation} 
where for simplicity we have suppressed the argument in $p_0,p_1$ and $q_0,q_1$.   This integral converges for $\tau_0>1$. Furthermore, the integrand has the following property \footnote{Had we not made the change of variables (\ref{xtoj}), the structure of the integrand would be much more complicated, including also $K_1(2x)$, and we couldn't have drawn the same conclusions so easily. One could have done an integration by parts, which takes $K_1(2x) \to K_0(2x)$ and produces derivatives of the other functions, but the computation would have been much more cumbersome.}:
\begin{equation}
\int_0^\infty j^{\tau_0-2} P(\log j^2/\epsilon)K_0(2j)dj = 0~~~ \rightarrow  ~~~P(\log j^2/\epsilon)=0
\end{equation} 
where $P(\log j^2/\epsilon)$ is a polynomial of any degree. Hence, assuming the leading twist operators are non-degenerate, to {\it any} loop order in perturbation theory (and since $q_0 \neq 0$), we obtain the following constraints
\begin{equation}
p_1 =0, ~~~q_1 = \frac{1}{2} q_0 p_0' \,,
\end{equation}  
provided $-\frac{\tau_0}{2}+\frac{1}{2}$ is negative. Considering higher powers $-\frac{\tau_0}{2}+\frac{1}{2} +n$ we get additional constraints, involving higher and higher orders in the expansions (\ref{largejexpansions}). For any given twist $\tau_0$, the powers $-\frac{\tau_0}{2}+\frac{1}{2} +n$ will become non-negative at some point, and the integral expression cannot be trusted any more. However, we can resort to the following trick: we can act on both sides of (\ref{OPEsmallu}) with the Casimir operator (\ref{casimir}). This will multiply the integrand by an overall factor $\frac{j^2}{\epsilon}$ and will allow us to explore one more order in the large $j$ expansion! Here it is important that acting with ${\cal D}$ on $h(\log u,v)$ will not spoil the analyticity properties. In this way, we can obtain constraints to arbitrarily high order in the large $j$ expansion. For instance, to the next half-integer power we obtain
\begin{equation}
\int_0^\infty j^{\tau_0-4} u^{\frac{1}{2} p_0} \left( 16 p_2 q_0 +16 q_3 -(q_0+8 q_2)p'_0 -8 q_0 p_2'+8 q_0 p_3 \log u \right)K_0(2j) dj = 0
\end{equation} 
At this order there is a priori complicated expression proportional to $K_1(2j)$, which vanish upon using the previous order constraints! As before, this implies

\begin{equation}
p_3 =0, ~~~q_3 = - p_2 q_0 +\frac{1}{16} (q_0+8 q_2)p'_0 +\frac{1}{2} q_0 p_2'- \frac{1}{2} q_0 p_3 \,,
\end{equation}  

and so on. Written in terms of the Casimir $J$, our findings can be summarized as follows:
\begin{itemize}
\item The expansion of $\gamma(J)$ for large $J$ contains only even powers of $1/J$.
\item The expansion of $\hat a(J)\left(1 - \frac{\sqrt{1+4 J^2}}{4J} \gamma'(J)\right)$ for large $J$ contains only even powers of $1/J$.
\end{itemize}
The first result is equivalent to the reciprocity principle for leading twist anomalous dimensions!  The second result is a new set of infinite conditions on structure constants. It can be written in terms of $\ell$ as:
\begin{equation}
\frac{\hat a(\ell)}{2+\gamma'(\ell)}~~~~ \textrm{has only even power when expanded in $1/J$.}
\end{equation}
Our results rely only on mild assumptions, and in particular are valid to any loop in perturbation theory.  

\bigskip

\noindent {\bf Comments on the super-symmetric case}

\bigskip

The results above rely on the assumption that leading twist operators with higher spins are non-degenerate. In general, this does not hold for  super-symmetric conformal field theories (SCFT), since the scalar operators of the form $\varphi \partial_{\mu_1} \cdots \partial_{\mu_\ell} \varphi$ will mix with operators of the schematic form $\bar \psi \gamma_{(\mu_1} \partial_{\mu_2} \cdots \partial_{\mu_\ell)} \psi$ and $F_{\nu ( \mu_1} \partial_{\mu_2} \cdots \partial_{\mu_{\ell-1}} F_{\mu_\ell) \nu}$. There are two ways to overcame this obstacle and apply our methods:
\begin{enumerate}
\item  A SCFT will have a global $R-$symmetry group. It is sometimes possible to project the correlator over a specific representation of the $R-$symmetry group such that only leading twist operators composed by scalars propagate as intermediate operators.  In this case one can apply our method straightforwardly. 

\item  For SCFT  one can organize the conformal partial wave expansion in terms of super-conformal blocks. In this case the sum runs over super-conformal primaries, which usually include only leading twist operators composed by scalars. In this case our method can again be applied, but the details will depend of the specific form of the super-conformal blocks. 
\end{enumerate}
In what follows we discuss how these options work for the case of ${\cal N}=4$ SCFT.

\subsection{Example: \texorpdfstring{$\mathcal{N}=4$}{} SYM}

A four-dimensional theory with abundance of perturbative results is ${\cal N}=4$ SYM. This theory has a $SU(4)$ R-symmetry group. Under this $R-$symmetry group scalars $\varphi^i$ transform in the ${\bf 6}$ representation, fermions in the ${\bf 4}$ and ${\bf \bar 4}$ and gauge bosons are singlets. The energy-momentum tensor lies in a half-BPS multiplet, whose superconformal primary is a scalar operator ${\cal O}$ of protected dimension $\Delta_{\cal O}=2$ and which transforms in the ${\bf 20}'$ representation of the R-symmetry group. When expanded in the $s-$channel, the correlator of four identical such operators will decompose into the various representations contained in ${\bf 20}' \times {\bf 20}'$.  The twist-two operators will contribute to the following representations:
\begin{eqnarray}
\mbox{Tr}\, \varphi \partial_{\mu_1} \cdots \partial_{\mu_\ell} \varphi ~ &\to& ~ {\bf 1} + {\bf 15} + {\bf 20'}\,,\\
\mbox{Tr}\, \bar \psi \gamma_{(\mu_1} \partial_{\mu_2} \cdots \partial_{\mu_\ell)} \psi ~ &\to& ~ {\bf 1} + {\bf 15}\,,\\
\mbox{Tr}\, F_{\nu ( \mu_1} \partial_{\mu_2} \cdots \partial_{\mu_{\ell-1}} F_{\mu_\ell) \nu} ~ &\to& ~ {\bf 1} \,.
\end{eqnarray}
If we project in the $ {\bf 20}'$,  only non-degenerate twist-two operators of the form $\mbox{Tr}\, \varphi^{( i} \partial^\ell \varphi^{j)}$ contribute. Their anomalous dimension, as well as their OPE coefficients, have been computed to three-loops in \cite{Eden:2012rr}. From these results we can compute the rescaled OPE coefficients. To two loops these take the form
\begin{align}\nonumber
\gamma(\ell)&=\lambda \gamma_1(\ell)+\lambda^2 \gamma_2(\ell)+\ldots \\
\hat{a}(\ell)&=1+ \lambda \,\hat{a}_1(\ell)+\lambda^2 \,\hat{a}_2(\ell)+\ldots
\end{align}
where we define the coupling constant $\lambda=\frac{g^2 N}{4 \pi^2}$ and
\begin{align*} 
\gamma_1(\ell)&=2 S_{1}(\ell)\,, \\
\gamma_2(\ell)&=-2 S_{-3}(\ell)-2 S_{-2}(\ell) S_{1}(\ell)-2 S_{1}(\ell) S_{2}(\ell)-S_{3}(\ell)+2 S_{-2,1}(\ell)\,,\\
\hat{a}_1(\ell)&=-S_{2}(\ell)\,,\\
\hat{a}_2(\ell)&=\frac{5}{2} S_{-4}(\ell)+S_{-2}^2(\ell)+2 S_{-3}(\ell)S_{1}(\ell)+\zeta_2 S_{1}^2(\ell)+S_{-2}(\ell)S_{2}(\ell)+S_{2}^2(\ell)\\
&+2S_{1}(\ell)S_{3}(\ell)+\frac{5}{2}S_{4}(\ell)-2S_{-3,1}(\ell)-S_{-2,2}(\ell)-2S_{1,3}(\ell)+3 \zeta_3 S_{1}(\ell) \,,
\end{align*}
where the harmonic sums are defined by
\begin{align*}
&S_a(\ell) = \sum_{m=1}^{\ell} \frac{1}{m^a},&S_{a,b,c,\ldots}(\ell) = \sum_{m=1}^{\ell} \frac{1}{m^a} S_{b,c,\ldots}(m)\,, \\
&S_{-a}(\ell) = \sum_{m=1}^{\ell} \frac{(-1)^m}{m^a},&S_{-a,b,c,\ldots}({\ell}) = \sum_{m=1}^{\ell} \frac{(-1)^m}{m^a} S_{b,c,\ldots}(m)\,.
\end{align*}
The three-loop results are quite cumbersome and not very illuminating. We have explicitly checked that these results (including three-loop) are consistent with the relations  of previous section, up to eight order in $1/J$.

Alternatively, we could organize our expansions in terms of super-conformal blocks. In this case the sum will run over super-conformal primaries. Among the twist two operators, only singlets made out of scalars, of the form $\mbox{Tr}\, \varphi^{i} \partial^\ell \varphi^{i}$ are super-conformal primaries. The corresponding super-conformal blocks have been worked out in \cite{Nirschl:2004pa}. It turns out they are simply given by the usual conformal blocks upon replacing $\Delta \to \Delta+4$. Our method will go through, after shifting the Casimir operator correspondingly:
\begin{equation}
\label{Jn4}
 \frac{1}{4} (2\ell+\gamma_\ell)(2\ell+2+\gamma_\ell)~\to~ \frac{1}{4} (2\ell+4+\gamma_\ell)(2\ell+6+\gamma_\ell)\,.
\end{equation}
On the other hand, as a consequence of superconformal symmetry, the anomalous dimensions of the singlet operators is given by the anomalous dimension of the operators in the ${\bf 20'}$, upon a shift $\ell \to \ell+2$. For instance, the one-loop anomalous dimension of the Konishi operator $\mbox{Tr}\, \varphi^{i} \varphi^{i}$ is proportional to $S_1(2)$. This shift in $\ell$  exactly accounts for the shift in (\ref{Jn4})! and our results apply. 

A similar study can be performed for the four-point correlator with external operators of larger dimensions. In particular, we can consider the case when the leading twist intermediate operators have twist three, e.~g.~two external operators with dimension $\Delta_1=\Delta_2=2$ and two with dimension $\Delta_3=\Delta_4=3$. The anomalous dimensions for this class of operators and their parity preserving properties were extensively studied in \cite{Beccaria:2008fi,Beccaria:2009eq,Velizhanin:2010cm}. In the case of twist-three operators the Casimir eigenvalue \eqref{casimir_pert} takes the form\footnote{Notice that it differs by a constant compared to the one used in \cite{Beccaria:2008fi}, however, it does not change the structure of the expansion.}
\begin{equation}
J^2_{\mbox{\tiny twist-3}}=\left(\ell+\frac{3}{2}+\frac{\gamma_\ell}{2}\right)\left(\ell+\frac{1}{2}+\frac{\gamma_\ell}{2}\right) 
\end{equation}
and our results easily apply. We have checked that indeed the anomalous dimension of twist-three operators available in the literature can be expanded using only even powers of $1/J_{\mbox{\tiny twist-3}}$.

\subsection{Comments on \texorpdfstring{$D \neq 4$}{} and applications to non-conformal theories}

Note that our derivation uses very little about the explicit form of conformal blocks. Namely, only their leading behavior as $u \rightarrow 0$. For identical external operators, it was shown in \cite{Dolan:2011dv} that the conformal blocks satisfy
\begin{equation}
G_{\Delta,\ell}(u,v) \sim u^{\sfrac{1}{2}(\Delta-\ell)}(1-v)^\ell ~_2F_1\left(\sfrac{1}{2}(\Delta+\ell),\sfrac{1}{2}(\Delta+\ell),\Delta+\ell;1-v\right),~~\textrm{as}~~u \to 0 \,,
\end{equation}
independently of the number of space-time dimensions. Hence, we expect our method to be applicable to CFT's in general dimensions. This opens up the possibility of applying our methods to a non-conformal theory, as follows.

As discussed in detail in \cite{Basso:2006nk} conformal symmetry implies the anomalous dimension of twist two operators with higher spin is a function of the conformal spin
\begin{equation}
\label{BK}
\gamma_\ell = f(\ell+\sfrac{1}{2} \gamma_\ell)\,.
\end{equation}
In this paper we have proven to all loops in perturbation theory that $\gamma_\ell$ admits an expansion in large $J^2 =(\ell+\gamma_\ell/2)(\ell+1+\gamma_\ell/2)$, involving only even powers of $J$. This proves reciprocity, which is equivalent to the parity preserving property of $f(\ell)$ stated in the introduction. 

As explained in \cite{Basso:2006nk}, if we were considering instead a gauge theory with non-vanishing beta-function, then the relation (\ref{BK}) will get modified, due to the breaking of conformal invariance. This breaking is scheme dependent. However, if we use dimensional regularization scheme (DREG) with $d=4-2\epsilon$, the beta function of the coupling is simply
\begin{equation}
\beta_\epsilon(g) = -2\epsilon +\beta(g)\,,
\end{equation}
where $\beta(g)$ is the beta function of the four dimensional theory. But then we note that $\beta_\epsilon(g)$ vanishes at $\epsilon_{cr} = \beta(g)/2$ and hence the gauge theory is conformal in $d_{cr}=4-2\epsilon_{cr}$ dimensions. Now we can apply our results after shifting the dimensions of the fundamental fields by $-\epsilon_{cr}$. Hence we expect the anomalous dimension, in the four dimensional non-conformal theory, to have an expansions in terms of the corrected Casimir
\begin{equation}
J^2_\beta =\left(\ell+\gamma_\ell/2-\beta/2\right)\left(\ell+1+\gamma_\ell/2-\beta/2\right)\,,
\end{equation}
which involves only even powers of $J_\beta$. For instance, it can be explicitly checked that this is the case for the two-loop quark transversity distribution in QCD \cite{Kumano:1997qp,Vogelsang:1997ak,Hayashigaki:1997dn}, as well as for the analogues in ${\cal N}=0,1,2$ SYM theories, whose expressions can be found in \cite{Belitsky:2005bu}.


\section{Non-perturbative CFT}

\subsection{Results}

One of the beautiful features of the conformal bootstrap program is that it also applies to CFT which do not possess a Lagrangian description. In this section we will see that the methods can be equally applied to CFT in the non-perturbative regime. 

As in the previous section, we consider the four-point function of four identical real scalar operators ${\cal O}$ of dimension $\Delta_{\cal O}$. Let us start by recalling the analysis of \cite{Fitzpatrick:2012yx,Komargodski:2012ek} . Let $\tau_{min}$ be the twist of the minimal twist operator appearing in the OPE of ${\cal O}$ with itself. Hence, for small values of $u$ we should have
\begin{equation}
{\cal G}(u,v) =1+ a_{\tau_{min},\ell_0} u^{\frac{\tau_{min}}{2}} (v-1)^{\ell_0} ~_2F_1(\ell_0 + \tau_{min}/2,\ell_0+\tau_{min}/2,2\ell_0+\tau_{min},1-v) +\ldots
\end{equation}
Crossing symmetry (\ref{crossing}) then implies a term of the form
\begin{align}
\label{crosstower}
{\cal G}(u,v) &= \frac{u^{\Delta_{\cal O}}}{v^{\Delta_{\cal O}}} \left( 1+ a_{\tau_{min},\ell_0} v^{\frac{\tau_{min}}{2}} (u-1)^{\ell_0} ~_2F_1(\ell_0 + \sfrac{1}{2}\tau_{min},\ell_0+\sfrac{1}{2}\tau_{min},2\ell_0+\tau_{min},1-u) +...\right) \nonumber \\
& =  \frac{u^{\Delta_{\cal O}}}{v^{\Delta_{\cal O}}} \left( 1+ a_{\tau_{min},\ell_0} v^{\frac{\tau_{min}}{2}} \left( \alpha \log u+ \beta+\ldots \right) +\ldots\right)\,,
\end{align}
where $\alpha,\beta$ are known expressions, but their form will be not important for us. We have suppressed higher powers in $u$ and $v$. As noted in
\cite{Fitzpatrick:2012yx,Komargodski:2012ek}, this implies the existence of a tower of operators of twist
\begin{equation}
\Delta_\ell -\ell= 2\Delta_{\cal O} +\gamma_\ell ,~~~~\gamma_\ell = - \frac{c}{\ell^{\tau_{min}}} +\ldots
\end{equation}
In order to apply our arguments, we note that given $\tau_{min}$, crossing symmetry together with the structure of conformal blocks, imply that the powers of $v$ that multiply $v^{\tau_{min}/2}$ in (\ref{crosstower}) are always integer. Hence, let us consider the contribution from that tower to the four point function, in the small $u$ limit. We obtain
\begin{equation}
\label{npsum}
\sum_{\ell=0,2,\ldots} a_\ell u^{\Delta_{\cal O}+\gamma_\ell/2}(1-v)^\ell F_{\Delta_{\cal O}+\ell+\gamma_\ell/2}(1-v) = \frac{u^{\Delta_{\cal O}}}{v^{\Delta_{\cal O}}} \left( 1+ a_{\tau_{min},\ell_0} v^{\frac{\tau_{min}}{2}} \left( \alpha \log u+ \beta+\ldots \right) +\ldots\right)
\end{equation}
The divergence in  $\frac{u^{\Delta_{\cal O}}}{v^{\Delta_{\cal O}}}$ fixes the behavior of $a_\ell$ at large $\ell$. We can simply take $a_\ell$ to be equal to $a_\ell^{(0)}$ in (\ref{atree}), upon replacing $\tau_{0} \rightarrow 2\Delta_{\cal O}$. In order to study the consequences of (\ref{npsum}) having only integer powers of $v$ times $v^{\tau_{min}/2-\Delta_{\cal O}}$ we can proceed as in the previous section. As before, we can define the rescaled OPE coefficients, exactly as in  (\ref{rescaledOPE}) upon replacing $\tau_{0} \rightarrow 2\Delta_{\cal O}$. As before, our results are better expressed in terms of the Casimir, which now takes the form
\begin{equation}\label{full_Casimir}
J^2 = (\ell+\Delta_{\cal O} +\gamma_\ell/2)(\ell+\Delta_{\cal O} +\gamma_\ell/2-1)\,.
\end{equation}
The leading behavior at large $J$ is fixed by the divergence $v^{\frac{\tau_{min}}{2}-\Delta_{\cal O}}$  to be
\begin{align}
\gamma_\ell &= \frac{c_1}{J^{\tau_{min}}}+\ldots\\
\hat a_\ell &= 1+  \frac{d_1}{J^{\tau_{min}}}+\ldots
\end{align}
where the coefficients $c_1,d_1$ can be fixed in terms of $\alpha,\beta$ in (\ref{crosstower}). What can we say about higher orders? The analysis depends on the precise value of $\tau_{min}$, for instance, the value of $2\tau_{min}$ versus $\tau_{min}+2$. Let us focus in the case $\tau_{min}=2$, which is the most common example. In this case we expect an expansion of the form 
\begin{align}\label{largejnonperturbative}
\gamma_\ell &= \frac{c_1}{J^2}+ \frac{c_2}{J^3}+ \frac{c_3}{J^4}+ \frac{c_4}{J^5}+\ldots \\
\hat a_\ell &= 1+  \frac{d_1}{J^2}+ \frac{d_2}{J^3}+ \frac{d_3}{J^4}+ \frac{d_4}{J^5}+\ldots 
\end{align}
Plugging these expansions into (\ref{npsum}), approximating the sums as we did in the previous section, and requiring half-integer divergent powers of $v$ to vanish, we find the constraints take exactly the same form as for the perturbative case:
\begin{itemize}
\item The expansion of $\gamma(J)$ for large $J$ contains only even powers of $1/J$.
\item The expansion of $\hat a(J)\left(1 - \frac{\sqrt{1+4 J^2}}{4J} \gamma'(J)\right)$ for large $J$ contains only even powers of $1/J$.
\end{itemize}
Note that these results allow for logarithmic dependence on $\log J$ for the expansion coefficients. These results can be trusted provided we don't get extra contributions from operators with twist close to $\tau_{min}$. The constraints for other values of $\tau_{min}$ take very much the same form. In many examples, as the ones seen below, $\gamma$ is proportional to an additional small parameter in which we are expanding only to first order. In this case we can expand in terms of the zeroth order Casimir $J_0$:
\begin{equation}
\label{J0}
J_0^2 =  (\ell+\Delta_{\cal O} )(\ell+\Delta_{\cal O} -1)\,.
\end{equation}

Note the crucial difference between the perturbative expansion \eqref{largejexpansions} and the non-perturbative one \eqref{largejnonperturbative}. These two expansions are not in contradiction, since they correspond to a priori different operators. For instance, in large $N$ gauge theories \eqref{largejexpansions} will correspond to single trace leading-twist operators, while \eqref{largejnonperturbative} will correspond to double trace operators, see \cite{Alday:2007mf}. Of course, as the coupling constant increases from zero to a finite value, the operator of leading twist should interpolate between a single trace and a double trace operator. To understand this interesting question is beyond the scope of the paper. 

\subsection{Examples}


\noindent {\bf Theories with gravity duals}

\medskip

The most well studied conformal field theory with gravity is ${\cal N}=4$ SYM in the large $N$ limit. In \cite{Hoffmann:2000dx} the four-point function of 2-2 dilaton scattering was considered. In this case, there is a tower of double trace operators of the form ${\cal O} \partial^\ell {\cal O}$ where ${\cal O}$ stands for the operator dual to the dilaton and has dimension four. The dimension of these double-trace operators was shown to be
\begin{equation}
\Delta_\ell - \ell = 8 - \frac{96}{N^2} \frac{1}{(\ell+1)(\ell+6)}\,.
\end{equation}
We see that the anomalous dimension behaves like $1/\ell^2$ for large values of the spin. This is consistent with the fact that the stress tensor is exchanged in the t-channel. It is easy to check that this result agrees with our relations. Indeed, setting $\Delta_{\cal O}=4$ in the zeroth order Casimir (\ref{J0}) we can obtain
\begin{equation}
 \frac{1}{(\ell+1)(\ell+6)}=\frac{1}{J_0^2-6}\,,
 \end{equation}
which contains only even powers of $1/J_0$.

\bigskip

\noindent {\bf Critical $O(N)$ models}

\medskip

Let us consider the four-point correlation functions of four spin fields $\sigma_i$. Among the intermediate states we have higher-spin states transforming in the singlet representation of the global $O(N)$ symmetry, of the form $\sigma_i \partial^\ell \sigma_i$, as well as states transforming in the symmetric traceless representation, of the form $\sigma_{(i} \partial^\ell \sigma_{j)}$. For the $O(N)$ critical model in $4-\epsilon$ dimensions, their anomalous dimensions have been computed to order $\epsilon^2$ in \cite{Wilson:1973jj}, with the result
\begin{eqnarray}
\label{ON4d}
\gamma_{\sigma_i \partial^\ell  \sigma_i} = 2\gamma_\sigma -\epsilon^2 \frac{3(N+2)}{(N+8)^2} \frac{1}{\ell (\ell +1)}\,,\\
\gamma_{\sigma_{(i} \partial^\ell  \sigma_{j)}} = 2\gamma_\sigma -\epsilon^2 \frac{(N+6)}{(N+8)^2} \frac{1}{\ell (\ell +1)}\,.
\end{eqnarray}
The leading power of the large $\ell $ behavior is governed by the presence of intermediate states of twist two (such as the stress tensor). Furthermore, to order $\epsilon^2$, the large $\ell $ expansion of these results is in perfect agreement with the relations above, where $\Delta_{\cal O}=1+\gamma_\sigma \approx 1$, is the dimension of the spin field in four dimensions. Indeed, written in terms of the zeroth order Casimir the anomalous dimensions behave exactly as $1/J_0^2$.

We can also consider the limit of large $N$ in $d$ dimensions \cite{Lang:1993ge,Lang:1992zw} and \cite{Derkachov:1997ch}\footnote{We thank the authors of \cite{Derkachov:1997ch} for pointing out this reference to us.}. At the leading order, for the operators in the symmetric traceless representation one obtains
\begin{eqnarray}
\gamma_{\sigma_{(i} \partial^\ell \sigma_{j)}}  - 2\gamma_\sigma  \sim \gamma_\sigma \frac{1}{(d+2\ell-4)(d+2\ell-2)}\,,
\end{eqnarray}
where $\gamma_\sigma \sim \frac{1}{N}$ and we have suppressed factors independent of $\ell$. In this case the intermediate operator with the lowest twist is $\sigma^2$, which has twist two and explains the leading power in the large $\ell$ expansion. Furthermore, one can explicitly check that the large $\ell$ expansion is in perfect agreement with our relation, where $\Delta_{\cal O}=\frac{1}{2}(d-2)$ is the dimension of the spin field in $d$ dimensions. Actually, when written in terms of the zeroth order Casimir this anomalous dimension is simply proportional to $1/J_0^2$.

At leading order for operators in the singlet representation we have a more interesting situation. Their anomalous dimension is
\begin{equation}
\label{onsym}
\gamma_{\sigma_i \partial^\ell \sigma_i} =\frac{8\gamma_\sigma}{(d+2\ell-4)(d+2\ell-2)}\left((d+\ell-2)(\ell-1)-\frac{\Gamma(d+1)\Gamma(\ell+1)}{4(d-1)\Gamma(d+\ell-3)} \right)\,.
\end{equation}
The large $\ell$ expansion now contains two superimposed series
\begin{align}
\label{onsinglet}
\gamma_{\sigma_i \partial^\ell \sigma_i} -2 \gamma_\sigma &=\gamma_\sigma\frac{d(2-d)}{2}\left( \frac{1}{\ell^2}+  \frac{3-d}{\ell^3}+ \frac{7+3/4 d(d-6)}{\ell^4}+\frac{(d-3)(d^2-6d+10)}{\ell^5}+\ldots\right) \nonumber\\
&+ \gamma_\sigma \frac{\Gamma(d+1)}{2-2d} \left(\frac{1}{\ell^{d-2}} -\frac{1}{2} \frac{(d-3)(d-2)}{\ell^{d-1}} +\ldots\right)\,.
\end{align}
The first series corresponds to the presence of $\sigma^2$, which has twist two. Written in terms of the zeroth order Casimir the whole series is again proportional to $1/J_0^2$. The leading behavior of the second tower is determined by the presence of conserved currents with twist $d-2$. Written in terms of the zeroth order Casimir it takes the form:
\begin{equation}
\frac{\Gamma\left(\frac{1}{2}\left(\sqrt{1+4 J_0^2}+5-d\right) \right)}{J_0^2 \Gamma\left(\frac{1}{2}\left(\sqrt{1+4 J_0^2}-3+d\right) \right)}\,,
\end{equation}
which can be seen to have an expansion of the form $1/J_0^{d-2}$ times even powers of $1/J_0$. 

The anomalous dimensions of operators in the symmetric traceless representation has been computed at order $1/N^2$ in \cite{Derkachov:1997ch}. 
The above analysis can be easily performed also in this case which is especially interesting since it involves the full Casimir eigenvalue instead of just its leading part $J_0$. The large $\ell $ expansion of anomalous dimension contains two series: one which corresponds to the presence of $\sigma^2$ and contains only even powers of the full Casimir eigenvalue $J$  \eqref{full_Casimir}; the second one, which corresponds to the presence of operators with twist $d-4$ and is of the form $1/J^{d-4}$ times even powers of $1/J$.

\section{Discussion}

Using CFT arguments we have derived an infinite number of constraints for the large spin expansion of the anomalous dimensions and structure constants of higher spin operators. In terms of the Casimir $J^2 =(\ell+\tau_0/2+\gamma_\ell/2)(\ell+\tau_0/2-1+\gamma_\ell/2)$, these constraints take the form:
\begin{itemize}
\item The expansion of $\gamma(J)$ for large $J$ contains only even powers of $1/J$.
\item The expansion of $\hat a(J)\left( 1 - \frac{\sqrt{1+4 J^2}}{4J} \gamma'(J) \right)$ for large $J$ contains only even powers of $1/J$.
\end{itemize}
Our arguments rely only on analyticity, unitarity, crossing-symmetry and the structure of the conformal partial wave expansion and apply to a large class of higher spin operators. For the case of conformal gauge theories our results provide a proof of the reciprocity principle to all orders in perturbation theory, but in addition provide a new "reciprocity" principle for structure constants. We have also argued, following \cite{Basso:2006nk}, that these results should extend also to non-conformal theories. 

Many comments are in order. Note that the perturbative proof did not use the full power of crossing symmetry. In \cite{Alday:2013cwa} the leading term in (\ref{smallvexpansion}) was considered, and crossing symmetry was used to derive the leading large spin behavior of the OPE coefficient, from that of the anomalous dimension. It would be interesting to use the full power of crossing symmetry to understand more about the structure of the solutions. In other words, one can see  \cite{Alday:2013cwa}  as solving the conformal bootstrap equation, in perturbation theory and at leading order in $u,v$. It would be very interesting to extend these results to higher orders.

As already mentioned, we have derived a new set of constraints on OPE coefficients. It would be interesting to test these constraints for examples in the literature, including non-conformal theories. On a more pragmatic spirit, it would be interesting to use our relations to constrain the form of possible OPE structures, as functions of the spin. This was certainly useful in the case of anomalous dimensions of leading twist operators in ${\cal N}=4$ SYM and even QCD.

Finally, a limitation of our method is that it gives definite results only for non-degenerate cases. There are very interesting examples involving degenerate twist operators, such as in ${\cal N}=1$ SCFT. It would be interesting to extend our results to this case. 

\section*{Acknowledgements}
We are grateful to Gregory Korchemsky and Sasha  Zhiboedov for comments on the manuscript. This work was supported by ERC STG grant 306260. L.F.A. is
a Wolfson Royal Society Research Merit Award holder.

\appendix

\section{Non-scalar correlators}
\label{appnonscalar}

In the body of the text we focused in the case of correlation functions of identical operators composed by scalars. In this appendix we relax this assumption and show that our conclusions remain unchanged. As in most of the body of the paper, we will work in four dimensions. 

The starting point for our perturbative discussion is the free theory correlator. When studying a scalar operator we have in mind external operators of the form ${\cal O} = \varphi^2$, (with or without trace). The correlator of four identical such operators was computed in a free-theory in \cite{Dolan:2000ut}. The small $u$ expansion takes the form
\begin{equation}
\label{smalluscalar}
{\cal G}(u,v) = 1+  u \,h_0^{scalar}(v)+\ldots,~~~~~~h_0^{scalar}(v)=\frac{1}{c} \left(  \frac{1}{v}+1 \right)\,,
\end{equation}
where $c$ is related to the central charge of the theory, but will play no role in our discussion. The leading twist operators in the OPE are twist-two operators of the form $\varphi \partial_{\mu_1} \cdots \partial_{\mu_\ell} \varphi$. From (\ref{smalluscalar}) we computed the corresponding OPE coefficients between two external operators and a twist-two operator. From these OPE coefficients we defined the rescaled OPE coefficients, which by construction are equal to 1 at tree level:
\begin{equation}
a_\ell^{scalar}=\frac{2 \Gamma \left(\ell+1 + \frac{\gamma_\ell}{2} \right)^2 \Gamma (\ell+2+ \frac{\gamma_\ell}{2} -1)}{\Gamma (\ell+1+ \frac{\gamma_\ell}{2})\Gamma (2 \ell+\gamma_\ell +1)} \hat a_\ell^{scalar}\,.
\end{equation}
Note that the prefactor is simply the tree-level OPE coefficients upon a rescaling $\ell \to \ell+ \gamma_\ell/2$. From this expression we have shown how to derive the reciprocity relations for $\gamma_\ell$ and $a_\ell^{scalar}$.

We can also consider a scalar operator which is a bilinear of fermions. Given a free fermion field $\psi$ we can consider ${\cal O} = \bar \psi \psi$. In four dimensions this operator has dimension 3. The correlator of four identical such operators was computed in a free-theory in \cite{Dolan:2000ut}. The small $u$ expansion takes the form
\begin{equation}
{\cal G}(u,v) = u \, h_0^{fermion}(v) +\ldots,~~~~~~h_0^{fermion}(v) = \frac{1}{c} \left( \frac{1}{v^2}-\frac{1}{v}-1+v \right) \,.
\end{equation}
The leading twist operators are now twist two operators of the form $\bar \psi \gamma_{(\mu_1} \partial_{\mu_2} \cdots \partial_{\mu_\ell)} \psi$. Again, one can compute the corresponding OPE coefficients and define the rescaled ones, which in this case take the form
\begin{equation}
\label{afermion}
a_\ell^{fermion}=\frac{\Gamma(\ell+1+\gamma_\ell/2)\Gamma(\ell+2+\gamma_\ell/2)}{\Gamma(2\ell+\gamma_\ell)}\hat a_\ell^{fermion}\,.
\end{equation}
Finally, we can also consider scalar operators of dimension four, of the form ${\cal O}=F_{\mu \nu} F^{\mu \nu}$. The four point function in the small u limit is then \cite{Dolan:2000ut}
\begin{equation}
{\cal G}(u,v) = u \, h_0^{vector}(v) +\ldots,~~~~~~h_0^{vector}(v) =\frac{1}{c} u \left( \frac{1}{v^3}-\frac{2}{v^2}-\frac{1}{v}-1+v \right) \,.
\end{equation}
Now the intermediate leading twist operators are $F_{\nu ( \mu_1} \partial_{\mu_2} \cdots \partial_{\mu_{\ell-1}} F_{\mu_\ell) \nu}$. The rescaled OPE coefficients take the form
\begin{equation}
\label{avector}
a_\ell^{vector}=(\ell+\gamma_\ell/2-1)\frac{\Gamma(\ell+1+\gamma_\ell/2)\Gamma(\ell+3+\gamma_\ell/2)}{4\Gamma(2\ell+\gamma_\ell)}\hat a_\ell^{vector}\,.
\end{equation}
The claim is that if we would redo all our computations with (\ref{afermion}) and (\ref{avector}) we would arrive to the same conclusions. The reason for this is very simple. One can simply notice the following relations
\begin{equation}
u \, h_0^{fermion}(v) = {\cal D} \, u \, h_0^{scalar}(v) ,~~~~~~u \, h_0^{vector}(v) = \left( \frac{1}{4} {\cal D}^2  -\frac{1}{2}{\cal D} \right) \, u \, h_0^{scalar}(v)\,. 
\end{equation}
This means that the ratios of the prefactors defining the rescaled OPE coefficients differ by even powers of the Casimir $J$ and all our conclusions go through. 

\section{A case with global symmetry}
So far we have considered leading twist "single-trace" operators of the form $\varphi \partial^\ell \varphi$, where $\varphi$ is a real scalar field. It is easy to see that $\ell=0,2,\ldots$ for conformal primary operators. Another interesting class of higher spin operators is of the form  $\varphi^\dagger \partial^\ell \varphi$, where $\varphi$ is now a scalar chiral field with a $U(1)$ charge. In this section we analyze the simplest four point function containing such operators as intermediate states, and show that our method applies also to that case. Let us consider the following correlator in four dimensions
\begin{equation}
\langle \varphi^\dagger(x_1) \varphi(x_2) \varphi^\dagger(x_3) \varphi(x_4)\rangle\,,
\end{equation}
where $\varphi$ has dimension one. At tree level this correlator reduces to 
\begin{equation}
{\cal G}^{(0)}(u,v) = 1+\frac{u}{v}\,.
\end{equation}
The leading twist operators correspond to the operators mentioned above. When performing the partial wave decomposition now we have to sum over odd as well as even spins:
\begin{equation}
\sum_{\ell=0,1,2,\ldots} a_\ell^0 u (v-1)^\ell F_{\ell+1}(1-v) =\frac{u}{v}\,.
\end{equation}
We find the simple result
\begin{equation}
a_\ell^0 = (-1)^\ell \frac{(\ell!)^2}{(2\ell)!}\,.
\end{equation}
Now we want to turn on the coupling constant $g$. The leading twist operators will acquire an anomalous dimension $\gamma_\ell$. An obstruction in applying the method of the body of the paper is that $\gamma_\ell$ and $\hat a_\ell$  are usually of the form
\begin{eqnarray}
\gamma_\ell = \gamma^{A}_\ell + \sigma(\ell) \gamma^{B}_\ell\,, \\
\hat a_\ell = \hat a^A_\ell +  \sigma(\ell) \hat a^B_\ell\,,
\end{eqnarray}
where $\sigma(\ell) = (-1)^\ell$. The presence of $\sigma(\ell)$ possesses several conceptual as well as technical obstacles. First of all, note that in order to state the reciprocity principle we require an analytic  continuation on $\ell$. While $\gamma_A$ and $\gamma_B$ are expected to have a precise analytic continuation, this is not the case with $\sigma(\ell)$. In applying our method, one can check that the presence of  $\sigma$ (or any odd power) will severely dump the sum over $\ell$ and sums containing $\sigma$ will not produce divergent terms \footnote{Namely, for any polynomial growing, the sum will not contribute a divergent term.}. Furthermore, if we try to perform the change of variables (\ref{xtoj}) we will generate an ugly $\sigma'(\ell)$, from the Jacobian. 

 Still, we can make progress. Let us show how this works for the simplest example of one loop. After introducing the integral representation for the hypergeometric function, we expand  the anomalous dimension and rescaled structure constants keeping only terms to order $g$. Then, we drop terms containing a $\sigma(\ell)$, since they will not contribute a divergent term (even after applying the Casimir operator an arbitrary number of times). At one loop we find that the piece $\gamma^{A}_\ell$  of the anomalous dimension contains only even powers in $1/J$, where $J^2 = \ell(\ell+1)$. For example, we could check this claim for the one-loop anomalous dimension of the twist-two Wilson operators in pure gluodynamics built from gauge fields of opposite helicity, see for instance \cite{Belitsky:2004sc}: 
 \begin{equation}
 \gamma_\ell = \psi(\ell+3) + \psi(\ell-1) - 2\psi(1) - \sigma(\ell) \frac{6}{(\ell+2)(\ell+1)\ell(\ell-1)}\,.
 \end{equation}
 We can explicitly check that indeed, the first piece admits an expansion in even powers of $1/(\ell(\ell+1))$. At higher loops it is convenient to consider two separate sums, one for spin odd and the other for spin even. As noted above, we would run into problems if we tried to use the full Casimir as a variable, so the best we can do is to introduce a change of variables, from $\ell \to J^2 =\ell(\ell+1)$. The structure of integrals is now much more complicated, but one get new constraints order by order in the $1/J$ expansion. Preliminary results lead us to conjecture that $f_+(\ell)+f_-(\ell)$ contains only even powers when expanded in $1/J$, where $f_\pm\left(\ell +\frac{1}{2}(\gamma^A\pm \gamma^B)\right) = \gamma^A\pm \gamma^B$.

\bibliographystyle{nb}
\bibliography{bibliography}

\end{document}